# DYNAMIC BANDWIDTH MANAGEMENT IN DISTRIBUTED VoD BASED ON THE USER CLASS USING AGENTS

**H S Guruprasad**
**Research Scholar, Dr MGR University**
**Asst Prof & HOD, Dept of ISE**
**BMS College of Engg, Bangalore**
hs_gurup@yahoo.com

**Dr. H D Maheshappa**
**Director**
**East Point Group of Institutions**
**Bangalore**
hdmappa@gmail.com

**Abstract - This paper proposes a dynamic bandwidth management algorithm in which more bandwidth is allocated for higher class users and also higher priority is given to the videos with higher popularity within a class using agent technology. The popularity and weight profile of the videos which is used for efficiently allocating bandwidth is periodically updated by a mobile agent. The proposed approach allocates more bandwidth for higher class users and gives higher priority for higher weight videos [popular videos] so that they can be served with high QoS, reduces the load on the central multimedia server and maximizes the channel utilization between the neighboring proxy servers and the central multimedia server and lower video rejection ratio. The simulation results prove the reduction of load on central multimedia server by load sharing among the neighboring proxy servers, maximum bandwidth utilization, and more bandwidth allocation for higher class users.**

## I. Introduction

Agents are autonomous programs which can understand an environment, take actions depending upon the current status of the environment using its knowledge base and also learn so as to act in the future. Autonomy, reactive, proactive and temporally continuous are mandatory properties of an agent. The other important properties are commutative, mobile, learning and dependable. These properties make an agent different from other programs. The agents can move around in a heterogeneous network to accomplish their assigned tasks. The mobile code should be independent of the platform so that it can execute at any remote host in a heterogeneous network [1, 2, 8, 10].

A video-on-demand system can be designed using any of the 3 major network configurations – centralized, networked and distributed. In a centralized system configuration, all the clients are connected to one central server which stores all the videos. All the client requests are satisfied by this central server. In a network system configuration, many video servers exist within the network. Each video server is connected to a small set of clients and this video server manages a subset of the videos. In a distributed system configuration, there is a central server which stores all the videos and smaller servers are located near the network edges. When a client requests a particular video, the video server responsible for the requests ensures continuous playback for the video [3].

In [5], Tay and Pang have proposed an algorithm called GWQ [Global Waiting Queue] which shares the load in a distributed VoD system and hence reduces the waiting time for the client requests. This load sharing algorithm balances the load between heavily loaded proxy servers and lightly loaded proxy servers in a distributed VoD. They assumed that videos are replicated in all the servers and videos are evenly required, which requires very large storage capacity in the individual servers. In [6], Sonia Gonzalez, Navarro, Zapata proposed a more realistic algorithm for load sharing in a distributed VoD system. Their algorithm maintains small waiting times using less storage capacity servers by allowing partial replication of videos. The percentage of replication is determined by the popularity of the videos. Proxy servers are widely used in multimedia networks to reduce the load on the central server and to serve the client requests faster.

In, [2], we had considered an architecture without PSG and a comparison was made with an architecture without neighbouring proxy servers. In this paper, we propose an efficient bandwidth allocation algorithm and VoD architecture for distributed VoD system which allocates higher bandwidth to the videos which have higher weights. The architecture consists of a Central Multimedia Server [CMS]. A set of local Proxy servers are connected together in the form of a ring to form a Local Proxy Server Group [PSG]. All the PSG's are connected to the CMS. All connections are made through fiber optic cables. The rest of the paper is organized as follows: Section 2 presents the proposed architecture, section 3 presents the proposed algorithm, Section 4 presents the simulation model, Section 5 presents the simulation results and discussion, Section 6 finally concludes the paper and further work.

## II. Proposed Architecture

In the proposed architecture shown in the Fig 1, a Central Multimedia Server [CMS] is connected to a group of proxy currently requested by its clients is stored in each proxy server.

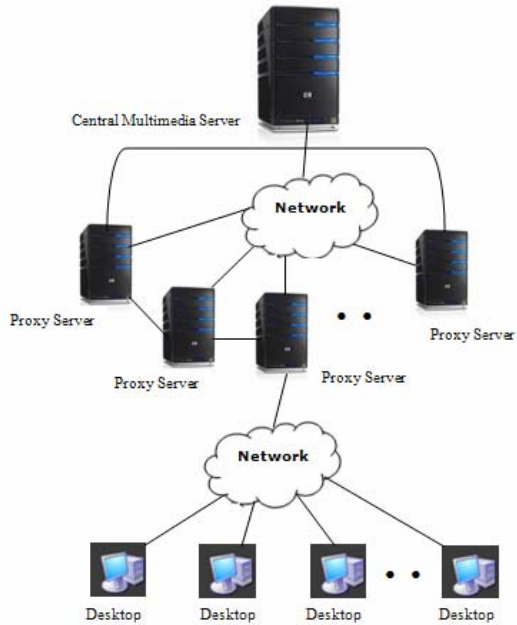

**Fig 1**

Consider n videos $v_1, v_2 \ldots \ldots v_n$. The mean arrival rates for the videos are $\lambda_1, \lambda_2 \ldots \ldots \lambda_n$ respectively. There are m server channels. The total arrival rate of all the videos is $\lambda = \sum_{i=1}^{m} \lambda_i$. The probability of receiving a user request for a video $v_i$ is given by $P_i = \lambda_i / \lambda$ for i=1, 2…..n.

There are 3 classes of customer's $c_1$, $c_2$ and $c_3$ and the profit associated with each class is $p_1$, $p_2$ and $p_3$ respectively. Let $k_1, k_2 \ldots \ldots k_n$ be the number of requests for the n videos $v_1, v_2 \ldots \ldots v_n$. Also, $k_i = k_{i1} + k_{i2} + k_{i3}$, where $k_{i1}$ is the number of requests of class 1, $k_{i2}$ is the number of requests of class 2 and $k_{i3}$ is the number of requests of class 3. Now, the weight associated with each video in class j is $w_i = k_{ij} * p_1$.

The CMS contains all the N number of videos. These N videos are categorized into most popular, secondary popular and least popular. Initially, few most popular, secondary popular and least popular videos are loaded to the proxy servers. Also there weights for these videos are appropriately assigned.

A single mobile agent is invoked by the CMS periodically and this mobile agent travels across the proxy servers and updates the video popularity and weight profiles at the proxy servers and the CMS.

When a request for a video arrives at a proxy server PS, one of the following 4 cases happens:

servers. All these proxy servers are connected through fiber optic cables in the form of a ring. Each proxy server is connected to a set of clients (users). The video content that is

-   The requested video is already present in the proxy server PS
-   The requested video is not present in the proxy server and the right neighbor proxy server[RPS], but is present in the left neighbor proxy[LPS] only
-   The requested video is not present in the proxy server and the left neighbor proxy server[LPS], but is present in the right neighbor proxy[RPS] only
-   The requested video is present in both LPS and RPS, but not in the proxy server PS
-   The requested video is not present in the proxy server, left neighbor proxy server[LPS] and right neighbor proxy server[RPS]

If the requested video is present in the proxy server, then the real time transmission of the video starts immediately from the proxy server to the client. If the requested video is not present in the proxy server, then the weight of the video is computed as explained above.

If the requested video is not present in PS and RPS, but is present in LPS, then the bandwidth for the requested video between PS and LPS is allocated as follows:

If maximum bandwidth required for the requested video in that class is available between PS and LPS, then maximum bandwidth is allocated. If not, minimum bandwidth for the video in that class is allocated if available between PS and LPS. If minimum bandwidth required for the video in that class is also not available for the requested video between PS and LPS, then we check if minimum bandwidth could be accumulated by deallocating excess allocated bandwidth than the minimum bandwidth starting from the bottom (i.e. least weighted video). This way excess bandwidth is taken starting from the lower weight videos in that class. If minimum bandwidth for the video in that class could be accumulated, then this minimum bandwidth is allocated to the requested video.

If bandwidth could not be allocated between PS and LPS, then bandwidth allocation is done between PS and CMS as explained above. If bandwidth could not be allocated between PS and CMS also, then the requested video is rejected.

If the requested video is not present in PS and LPS, but is present in RPS, then the bandwidth for the requested video between PS and RPS is allocated as explained above. If bandwidth could not be allocated between PS and RPS, then bandwidth allocation is done between PS and CMS as explained above. If bandwidth could not be allocated between PS and CMS also, then the requested video is rejected.

If the requested video is not present in PS, but is present in both LPS and RPS, then we check for the free bandwidth available between PS-LPS and PS-RPS. If free bandwidth available between PS and LPS is more than the free bandwidth available between PS and RPS, then bandwidth allocation is done between PS-LPS, otherwise bandwidth allocation is done between PS-RPS. If bandwidth could not be allocated between PS-RPS and PS-LPS, then bandwidth allocation is done between PS and CMS as explained above. If bandwidth could not be allocated between PS and CMS also, then the requested video is rejected.

### III. Proposed Algorithm

[Nomenclature

    PS: proxy server

    LPS: Left neighbor proxy server

    RPS: Right neighbor proxy server

    BW: Bandwidth

    BWAvail (x, y): Bandwidth available between x and y

    $MaxBW_i$: Maximum Bandwidth for the video in class i

    $MinBW_i$: Minimum Bandwidth for the video in class i]

When a request for a video arrives at a particular time t by a user of class i, do the following:

if the requested video is present in PS

    start streaming the video from PS

else

    dynamic bandwidth allocation is done according to the algorithm DynamicBand

    if bandwidth is allocated

    then

        the video is downloaded and stored at PS and streamed to the requested client

    else

        the request is rejected

Algorithm DynamicBand

begin

    if the requested video is present in LPS only

    then begin

        call BA (LPS, PS, i)

        if bandwidth is not allocated

            call BA (CMS, PS, i)

    end

    if the requested video is present in RPS only

    then begin

        call BA (RPS, PS , i)

        if bandwidth is not allocated

            call BA (CMS, PS, i)

    end

    if the requested video is present in both LPS and RPS

    then begin

        if (BWAvail (LPS, PS)>BWAvail (RPS, PS))

        then begin

            call BA (LPS, PS, i)

            if bandwidth is not allocated

                call BA (CMS, PS, i)

        end

        else begin

            call BA (RPS, PS, i)

            if bandwidth is not allocated

                call BA (CMS, PS, i)

        end

    end

    if the requested video is not present in LPS and RPS

    then

        call BA (CMS, PS, i)

end

Algorithm BA(X, PS, i)

begin

    if $MaxBW_i$ required for the video is available between X and PS

    then

        allocate $MaxBW_i$ for the video

    else

        if $MinBW_i$ required for the video is available between X and PS

        then

            allocate $MinBW_i$ for the video

        else

            find out if $MinBW_i$ required for the requested video could be accumulated by deallocating excess BW than the MinBW starting from the bottom (i.e. least weighted video in class i)

        if $MinBW_i$ could be accumulated

        then

            allocate $MinBW_i$ required for the requested video

        else

            BW is not allocated for the requested video

end

### IV. Simulation Model

    The simulation model consists of a single central multimedia server and a few proxy server groups. The PSG consist of a few proxy servers. The parameters considered for simulation are as follows:

| Parameter | values |
|---|---|
| Number of proxy servers | 6 |
| Number of videos[NOV] | 480 |
| Bandwidth between PS-LPS, PS-RPS and PS-CMS | 300MB |
| Max Bandwidth for the videos in class 1 | 24MB to 29MB |
| Min Bandwidth for the videos in class 1 | 8MB to 11MB |
| Max Bandwidth for the videos in class 2 | 18MB to 23MB |
| Min Bandwidth for the videos in class 2 | 6MB to 8MB |
| Max Bandwidth for the videos in class 3 | 12MB to 17MB |
| Min Bandwidth for the videos in class 3 | 4MB to 6MB |
| No. of most popular videos | NOV/4 =120 |
| No. of secondary popular videos | NOV/4 = 120 |

| No. of least popular videos | NOV/2 =240 |
|---|---|
| No. of most popular videos initially loaded to PS | 40 |
| No. of secondary popular videos initially loaded to PS | 40 |
| No. of least popular videos initially loaded to PS | 80 |

The performance parameters are load sharing among the proxy servers, more bandwidth allocation for the videos having more weights, video rejection ratio and the bandwidth utilization between PS-LPS, PS-RPS, PS-CMS.

## V.  Results and discussion

The results presented are an average of several simulations conducted on the model. Each simulation is carried out for 10000 seconds.

It is assumed that the video requests for the most popular, secondary popular and least popular videos are 50%, 35% and 15% respectively. Also, it is assumed that 20% of the requests are by class 1 users, 30% of the requests are by class 2 users and 50% of the requests are by class 3 users. The size of the videos are assumed to be quite large.

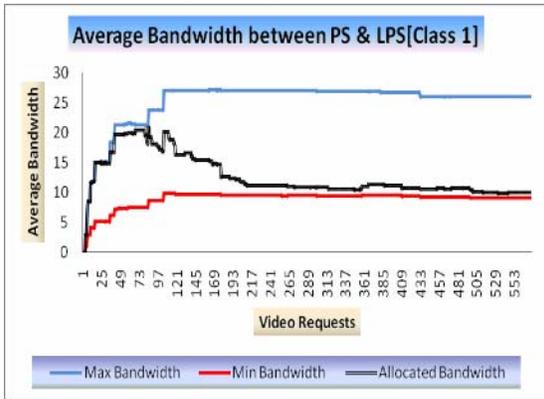

Fig 2

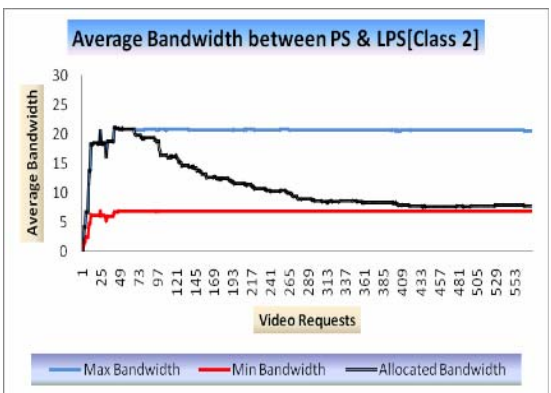

Fig 3

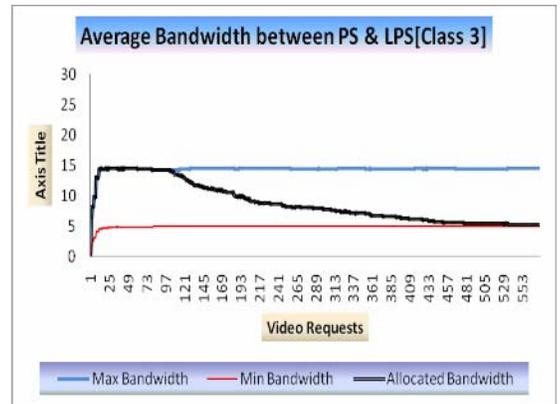

Fig 4

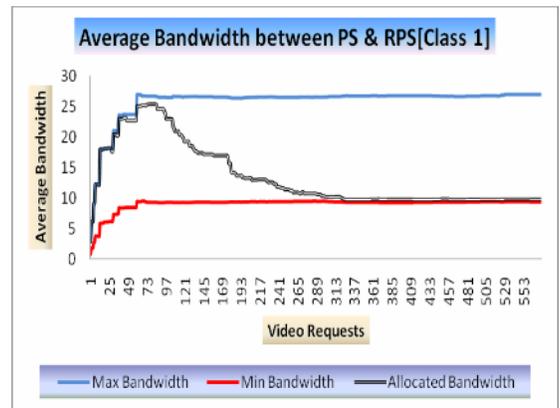

Fig 5

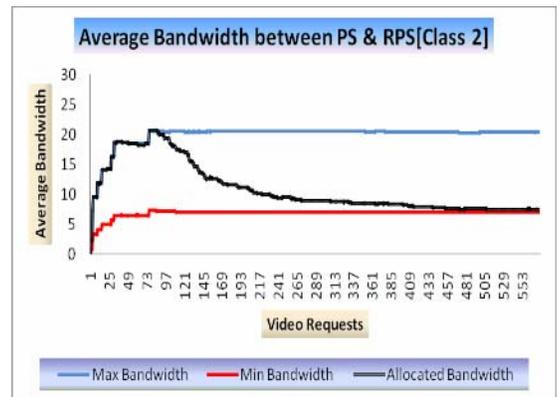

Fig 6

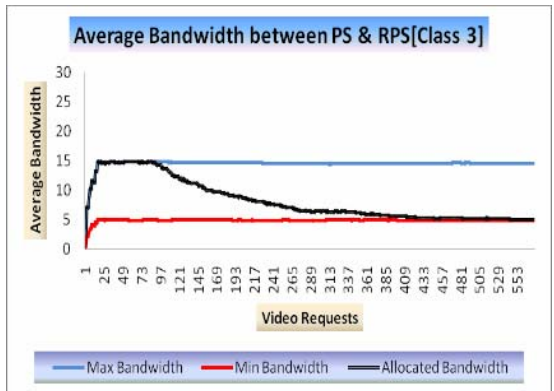

Fig 7

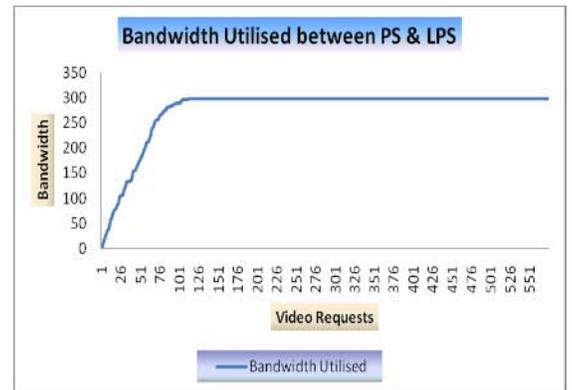

Fig 11

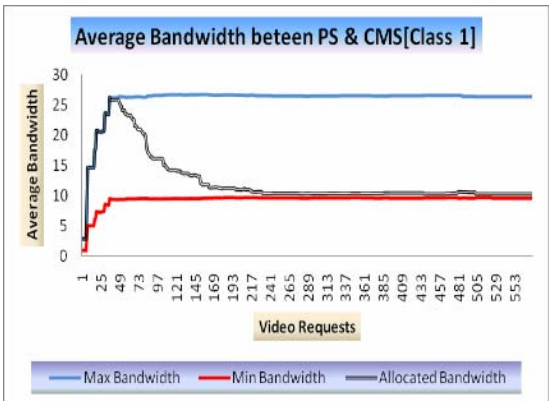

Fig 8

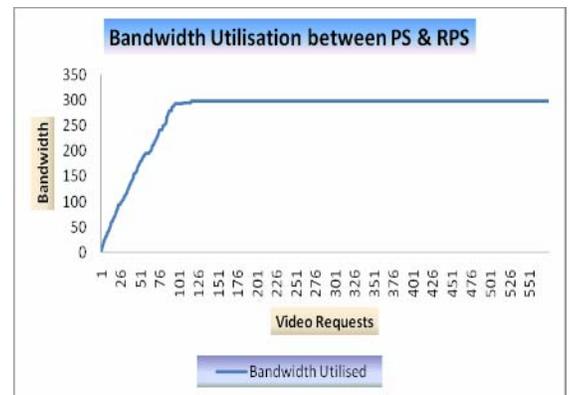

Fig 12

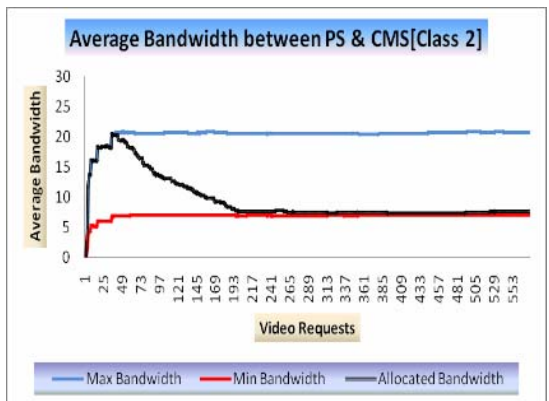

Fig 9

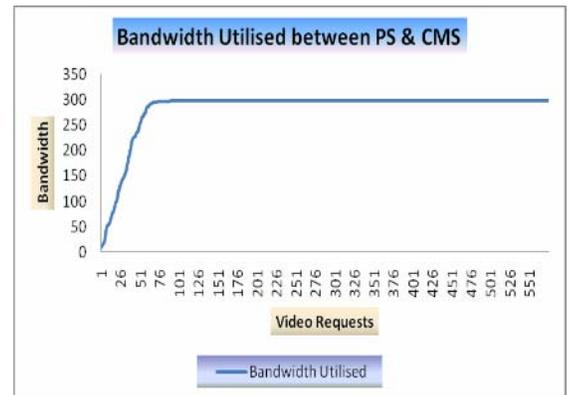

Fig 13

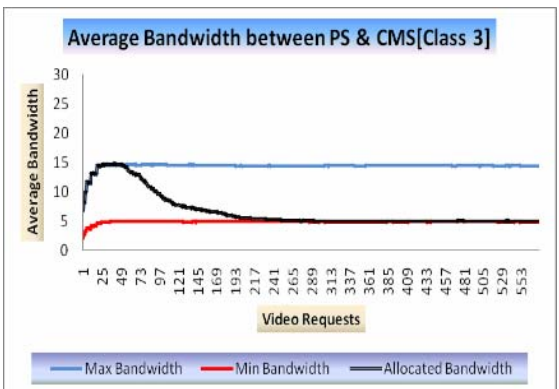

Fig 10

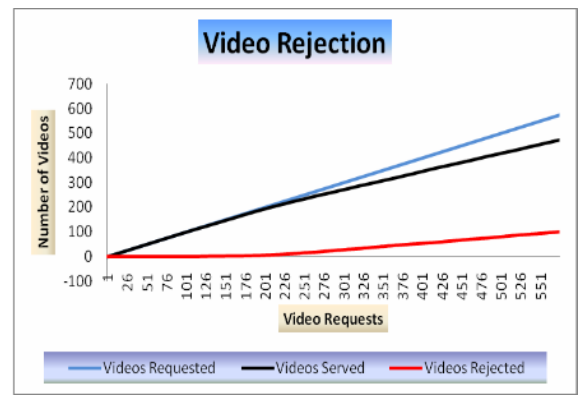

Fig 14

Fig 2, Fig 3 and Fig 4 shows the average maximum bandwidth, average minimum bandwidth and the average allocated bandwidth for all the videos being downloaded from LPS for class 1, class 2 and class 3 respectively. Initially maximum bandwidth is allocated to the videos downloaded from LPS in all the 3 classes. Later, when the number of videos being downloaded from LPS increases, the excess bandwidth of the lower weight videos in that class being downloaded will be taken back to allocate for new videos. Thus more bandwidth will be assigned to the more weight videos than the lesser weight videos. When the number of videos still increases, then the average bandwidth allocated still decreases.

Fig 5, Fig 6 and Fig 7 shows the average maximum bandwidth, average minimum bandwidth and the average allocated bandwidth for all the videos being downloaded from RPS for class 1, class 2 and class 3 respectively. Initially maximum bandwidth is allocated to the videos downloaded from RPS in all the 3 classes. Later, when the number of videos being downloaded from RPS increases, the excess bandwidth of the lower weight videos in that class being downloaded will be taken back to allocate for new videos. Thus more bandwidth will be assigned to the more weight videos than the lesser weight videos. When the number of videos still increases, then the average bandwidth allocated still decreases.

Fig 8, Fig 9 and Fig 10 shows the average maximum bandwidth, average minimum bandwidth and the average allocated bandwidth for all the videos being downloaded from CMS for class 1, class 2 and class 3 respectively. Initially maximum bandwidth is allocated to the videos downloaded from CMS in all the 3 classes. Later, when the number of videos being downloaded from CMS increases, the excess bandwidth of the lower weight videos in that class being downloaded will be taken back to allocate for new videos. Thus more bandwidth will be assigned to the more weight videos than the lesser weight videos. When the number of videos still increases, then the average bandwidth allocated still decreases.

Fig 11, Fig 12 and Fig 13 shows the bandwidth utilisation between the PS-LPS, PS-RPS and PS-CMS. The bandwidth utilisation is almost maximum as shown in the figures. Thus, the bandwidth between PS-LPS, PS-RPS and PS-CMS is effienly used.

Fig 14 shows the number of videos requested, videos rejected without PSG and the videos rejected. The number of rejections is quite low which majorly happens when the video is not found in LPS and RPS also there is no free bandwidth between PS and CMS.

## VI.   Conclusion

In this paper, we have concentrated on the dynamic bandwidth management among the proxy servers by considering the user class and the popularity of the videos using agents. The simulation shows promising results. The algorithm always uses maximum bandwidth between the neighboring proxy servers and the central multimedia server by allocating more bandwidth to the higher class users and also to the popular videos in the class so that they are served with high QoS. Further work is being carried out to investigate dynamic bandwidth management by considering a local proxy server group.

AUTHORS PROFILE

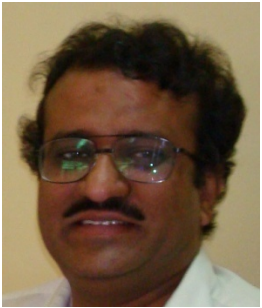

H S Guruprasad is an Assistant Professor and Head of the Department of Information Science & Engineering, BMS College of Engineering, Bangalore, India. He is an Engineering Graduate from Mysore University and did his Post Graduation at BITS, Pilani, India in 1995. He has vast experience in teaching and has guided many post graduate students. His research interests include multimedia Communications, distributed systems, computer networks and agent technology.

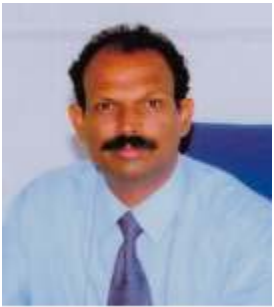

Dr. H.D. Maheshappa Graduated in Electronics & Communication Engineering from University of Mysore, India in 1983. He has a Post Graduation in Industrial Electronics from University of Mysore, India in 1987. He holds a Doctoral Degree in Engineering from Indian Institute of Science, Bangalore, India, since 2001. He is specialized in Electrical contacts, Micro contacts, Signal integrity interconnects etc. His special interests in research are Bandwidth Utilization in Computer Networks. He has been teaching engineering for UG & P G for last 25 years. He served various engineering colleges as a teacher and at present he working as director, East point group of institutions, Bangalore, India. He has more than 35 research papers in various National and International Journals & Conferences. He is a member of IEEE, ISTE, CSI & ISOI. He is a member of Doctoral Committee of Coventry University UK. He has been a Reviewer of many Text Books for the publishers McGraw-Hill Education (India) Pvt., Ltd, Chaired Technical Sessions, National Conferences and also has served on the advisory and technical national conferences.